\documentclass{iopart}
\usepackage{iopams}
\usepackage{graphicx}
\usepackage{units}
\usepackage{subfig}
\usepackage{ulem}
\usepackage{color}
\usepackage[latin1]{inputenc}
\usepackage{cite}
\begin{document}

\title{Optical Absorption Measurement at \unit[1550]{nm} on a Highly-Reflective Si/SiO$_2$ Coating Stack}

\author{Jessica Steinlechner, Alexander Khalaidovski and\\ Roman Schnabel}

\address{Institut f\"ur Gravitationsphysik,
Leibniz Universit\"at Hannover and Max-Planck-Institut f\"ur
Gravitationsphysik (Albert-Einstein-Institut),\\  Callinstr. 38,
30167 Hannover, Germany}

\ead{roman.schnabel@aei.mpg.de}

\date{\today}
%
%
\begin{abstract}

Future laser-interferometric gravitational wave detectors (GWDs) will potentially employ test mass mirrors from crystalline silicon and a laser wavelength of 1550\,nm, which corresponds to a photon energy below the silicon bandgap. Silicon might also be an attractive high-refractive index material for the dielectric mirror coatings. Films of amorphous silicon (a-Si), however, have been found to be significantly more absorptive at 1550\,nm than crystalline silicon (c-Si).
Here, we investigate the optical absorption of a Si/SiO$_2$ dielectric coating produced with the ion plating technique. The ion plating technique is distinct from the standard state-of-the-art ion beam sputtering technique since it uses a higher processing temperature of about 250$^\circ$C, higher particle energies, and generally results in higher refractive indices of the deposited films.  
Our coating stack was fabricated for a reflectivity of $R=99.95\,\%$ for s-polarized light at 1550\,nm and for an angle of incidence of 44$^\circ$. We used the photothermal self-phase modulation technique to measure the coating absorption in s-polarization and p-polarization. We obtained $\alpha^{\rm coat}_{s}=\unit[(1035 \pm 42)]{ppm}$ and $\alpha^{\rm coat}_{p}=\unit[(1428 \pm 97)]{ppm}$. These results correspond to an absorption coefficient which is lower than literature values for a-Si which vary from $\unit[100]{/cm}$ up to $\unit[2000]{/cm}$. It is, however, still orders of magnitude higher than expected for c-Si and thus still too high for GWD applications.

\end{abstract}

%
%
\section{Introduction}

Highly-reflective (HR) dielectric mirror coatings are essential for interferometric gravitational wave detectors (GWDs). Optical coatings with low optical absorption and low mechanical loss are required~\cite{coating_loss}. It is expected that mechanical loss in optical coatings will result in a significant thermal noise source in GWDs of the 2nd generation, which are currently under construction~\cite{thermal_noise}. The 2nd generation of GWDs will operate at the well-established laser wavelength of \unit[1064]{nm} and will use well-established dielectric optical coatings made of alternating layers of silica (SiO$_2$) and tantala (Ta$_2$O$_5$). 
The test mass mirror HR coatings  for the 1st generation LIGO GWDs~\cite{ligo_ca} had a design transmission of \unit[5]{ppm}~\cite{aligo_ca} and an absorption requirement of $\leq\unit[1]{ppm}$. For the 2nd generation Advanced LIGO GWDs the optical absorption needs to be less than \unit[0.5]{ppm}~\cite{coating_loss, aligo_ca}. While coatings with the specified absorption requirement were already produced~\cite{coating_loss}, the mechanical loss realized is still above the requirement of $\phi=5\times 10^{-5}$~\cite{aligo_ca}, at which tantala contributes most to the coating's mechanical loss~\cite{thermal_noise}. Doping the tantala with titanium reduces the loss by nearly a factor of two, but slightly increases the coating absorption~\cite{coating_loss}. In addition, titanium doped tantala shows a loss peak at low temperatures~\cite{tantala}, where some of the 3rd generation GWDs will operate~\cite{hild2}. Therefore various materials are under investigation to replace tantala within dielectric coatings~\cite{schnabel2010}.

Because of its high mechanical Q-factor at cryogenic temperatures~\cite{Nawrodt2008,Guigan1978}, crystalline silicon (c-Si) is considered as a test-mass material~\cite{schnabel2010} for future GWDs. c-Si shows a high optical absorption at \unit[1064]{nm}. Since the absorption decreases rapidly toward higher wavelengths~\cite{Keevers1995}, the use of c-Si test masses is planned at a wavelength of \unit[1550]{nm},  which corresponds to a photon energy below the silicon band gap. 
Silicon might also be an interesting candidate for a coating material in view of replacing tantala. The high index of refraction of Si of $n_{\rm Si}=3.48$ at \unit[1550]{nm}~\cite{frey06} significantly reduces the number of single layers required to achieve high reflectivities. It also reduces the geometrical thickness of the coating stack, since a single layer has a constant optical thickness of about a quarter of the wavelength. The optical as well as the mechanical loss of the Si films, however, need to be less than the values for tantala. Dielectric coating stacks usually contain amorphous material, and amorphous Si (a-Si) is known to be rather absorptive at \unit[1550]{nm}~\cite{Chittick70,Brodsky69,Loveland73}. In~\cite{Brodsky69} it is reported that annealing of a-Si reduces the absorption coefficient of the material. The a-Si in~\cite{Brodsky69} was produced with radio-frequency sputtering.

In this work we present optical absorption measurements at \unit[1550]{nm} on a highly-reflective Si/SiO$_2$ coating produced with the ion plating technique. The coating was produced with high particle energies between 30\,eV and 50\,eV, a high processing temperature of about $\unit[250]{^\circ C}$, and was finally annealed at $\unit[400]{^\circ C}$. The coating was fabricated by {\it{Tafelmaier D{\"u}nnschicht-Technik GmbH}}~\cite{tafelmaier} and optimized for a reflectivity of $R=99.95\,\%$ at \unit[1550]{nm} in s-polarization and an angle of incidence (AOI) of $44^\circ$. The ion plating technique is distinct from today's state-of-the-art ion beam sputtering technique since it uses a higher processing temperature, higher particle energies, and generally results in higher refractive indices of the deposited films. The optical absorption of Si-films produced with this technique is not known. Our absorption measurement was performed in a three-mirror ring-cavity setup (see Fig.~\ref{fig:experimental_setup}\,(a.)) in s- and p-polarization using the photothermal self-phase modulation technique~\cite{SHG,Stei12}.  
The actual design of the coating stack was not at our disposal. Therefore, we used a model coating stack~\cite{lightmachinery} that matched our measurement results for the reflectivities in s- and p-polarization.  From this model we calculated the absorption coefficient of the Si layers taking into account an effective penetration depth of the laser beam.

%
%
\section{Experimental setup and results}

\begin{figure}
  \centering
  \includegraphics[width=11cm]{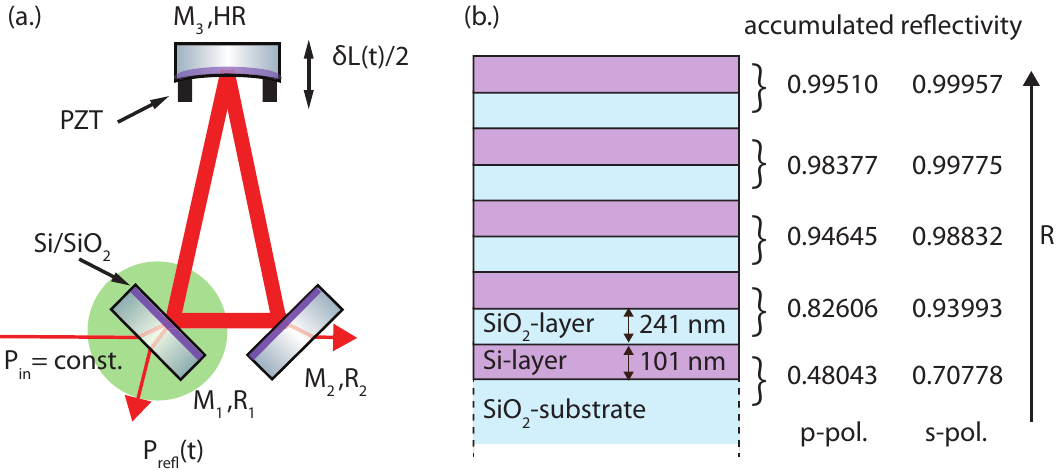}
  \caption{\footnotesize (a.) Experimental setup of the three-mirror ring-cavity with mirror power reflectivities $R_1 = R_2 = 99.947\,\%$ (measured in s-polarization) and $R_3 > R_{1,2} $ being close to unity. The cavity length was changed by a piezoactuator (PZT) mounted between the cavity spacer and mirror $M_3$. The absorption of the coating of in-coupling mirror $M_1$ (green) was measured in s-polarization and p-polarization at an angle of incidence of $44^\circ$. (b.) Schematic of our dielectric Si/SiO$_2$-coating model that we used to transfer our measurement results to absorption coefficients of the ion plating deposited silicon. By matching the reflectivities to our measurement results, the thickness of the single layers and the reflectivities for the double layers were calculated using~\cite{lightmachinery}.\\}
  \label{fig:experimental_setup}
\end{figure}

The technique of photothermal self-phase modulation (PSM) allows coating absorption measurements, if the substrate that carries the mirror coating is used as the incoupling mirror of an optical cavity~\cite{SHG,Stei12}. For our experiment we used a three-mirror ring-cavity formed by the plane in-coupling mirror $M_1$ and end-mirror $M_2$ together with the concave highly-reflective (HR) coated mirror $M_3$ (see Fig.~\ref{fig:experimental_setup}\,(a.)). All three mirrors had substrates from Corning 7980 glass~\cite{corning}. All coatings were manufactured by Tafelmaier~\cite{tafelmaier} using the ion plating technique. The coatings of $M_1$ and $M_2$ were produced in the same coating run and designed for a reflectivity of $R=\unit[(99.95 +0.01/-0.03)]{\%}$ at an angle of incidence (AOI) of $44^\circ$ for a wavelength of \unit[1550]{nm} and s-polarization. The HR coating of $M_3$ had negligible transmission compared to the transmissions of $M_1$ and $M_2$. The mirror substrate parameters and cavity geometry parameters are presented in Tab.~\ref{tab:parameters}.

\begin{table}
 \centering
  \caption{\footnotesize Material and geometric parameters of the Corning 7980 mirror substrates
  and our cavity, respectively, as used to simulate the self-phase modulation and to derive the coating absorption.}
 \label{tab:parameters}
{\footnotesize
 \begin{tabular}{llll}
 \hline
        Material parameters			 					& 																								&\multicolumn{2}{l}{Cavity geometry parameters}\\
        \multicolumn{4}{l}{(at $\unit[1064]{nm}$ and room temperature)}																													\\
 \hline
 				index of refraction $n$						& $1.48$ \cite{leviton06}													&round-trip length $L$ 						& \unit[42]{cm}\\
        thermal refr. coeff. d$n$/d$T$		& $\unit[8.45\cdot 10^{-6}]{/K}$\cite{leviton06} 	&beam waist (radius) $w_0$ 	& \unit[448]{$\mu$m}\\
        specific heat $c$ 								& $\unit[770]{J/(kg\,K)}$ \cite{val}							&mirror length $D$ 								& \unit[6.35]{mm}\\
        density $\rho$ 										& $\unit[2201]{kg/m^3}$ \cite{corning}						&mirror radius $r$ 								& \unit[12]{mm}\\
        thermal expansion $a_{\rm th}$ 		& $\unit[0.52\cdot 10^{-6}/]{K}$ \cite{corning}		&AOI															& $44^\circ$\\
        thermal conductivity $k_{\rm th}$ & $\unit[1.3]{W/(m\,K)}$ \cite{corning}						&																	&\\
 \hline
 \end{tabular}
}
 \hspace{0.5 cm}
 \vspace{0.4 cm}
 \end{table}

The three mirrors forming the cavity were glued to an aluminum spacer. In the course of our PSM technique, the cavity length was scanned and the shapes of the cavity resonance peaks were recorded. A piezoactuator (PZT) between $M_3$ and the spacer was used to modulate the cavity round-trip length. To minimize a potential nonlinear motion of the PZT the cavity length was scanned only in a small range around a cavity resonance of approximately \unit[10]{\%} of a free spectral range. The modulation voltage was constant for all measurements. 
To compensate for an remaining nonlinearity of the PZT motion as well as to compensate for a potential PZT hysteresis the actual mirror motion for each modulation frequency and ramp side (expansion or contraction of the PZT) was calibrated using phase-modulation side-bands imprinted on the laser signal via an electro optical modulator (EOM) before entering the cavity. The resonance peaks for the absorption measurements were detected in reflection of $M_1$ with a photo diode (PD).

In this setup, the PSM technique provides the absorption of the coating of the in-coupling mirror $M_1$. While the intra-cavity field builds up, the coating absorbs light. The resulting heat is transferred to the substrate of the incoupling mirror and the in-coupled beam receives a phase shift due to the change of the optical path length through $M_1$ resulting in a broadening and narrowing of the cavity resonance peak. At the same time absorption in the coatings of all three mirrors result in thermal expansion of the mirrors, which shortens the cavity round-trip length. The measured deformation of the cavity resonance peaks is, however, clearly dominated by the first effect.

\begin{figure}
	\centering
    \includegraphics[width=11cm]{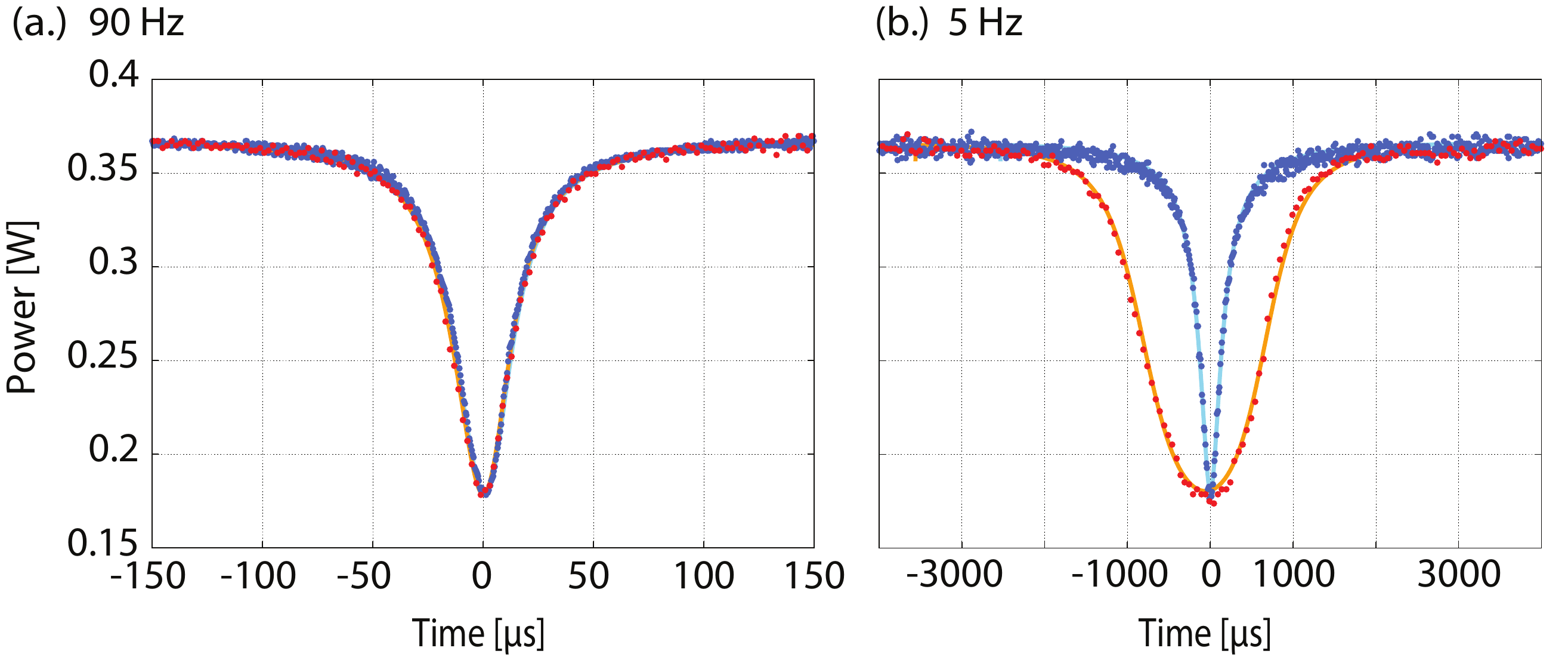}
    \caption{\footnotesize Cavity resonance peaks in s-pol for scan frequencies of \unit[90]{Hz} and \unit[5]{Hz} using the same scan amplitude and input light power measured on the reflected light. While at \unit[90]{Hz} (a.) no thermal effect occurs and the peaks are identical for lengthening and shortening of the cavity, at \unit[5]{Hz} (b.) for an external lengthening of the cavity the broad peak (red dots: measurement, orange line: simulation) forms, for an external shortening the narrow peak (dark-blue dots: measurement, light-blue line: simulation) forms. From this thermal effect the coating absorption was derived.}
    \label{fig:SiSiO2_Peaks}
\end{figure}

\begin{figure}
  \centering
  \includegraphics[width=11cm]{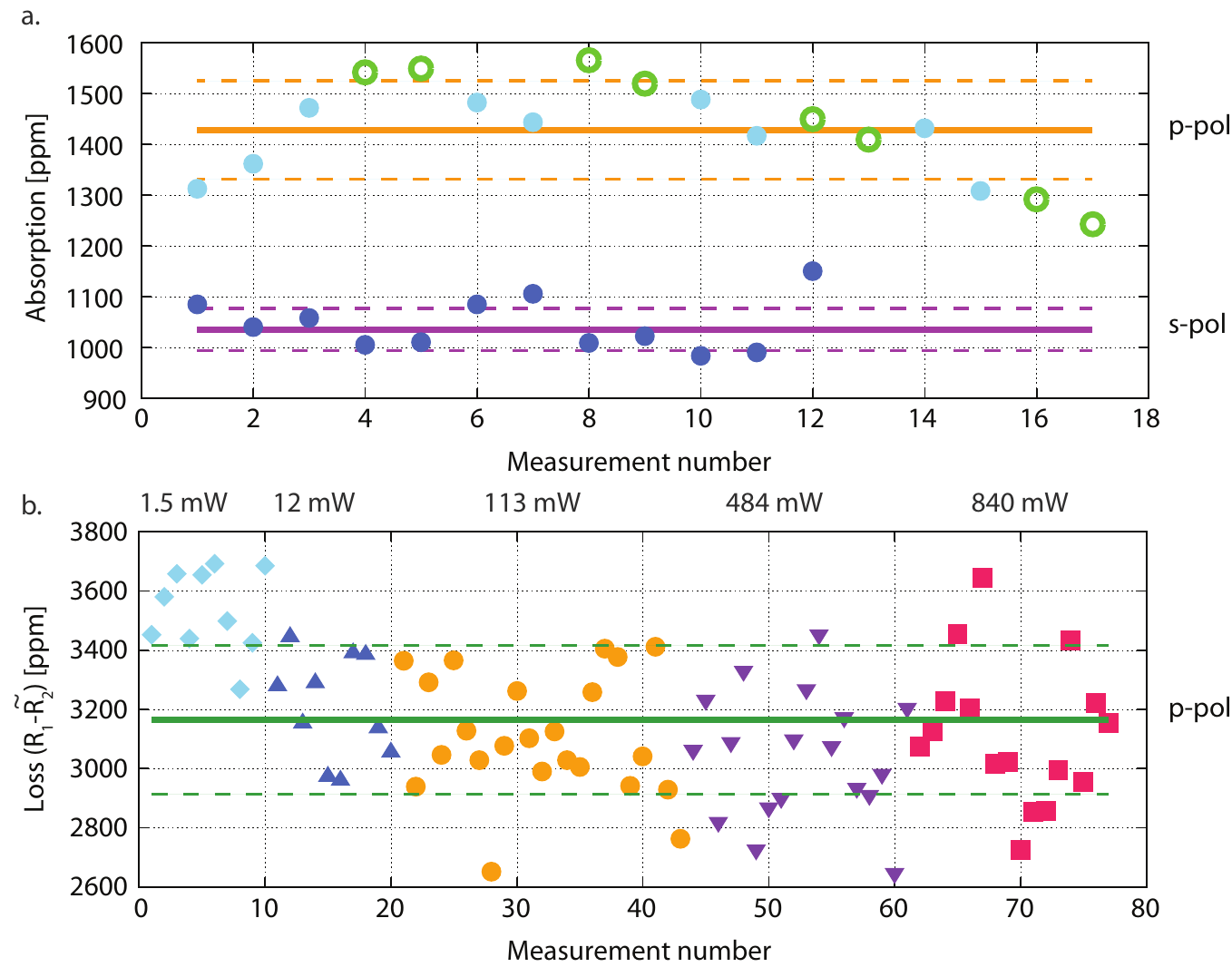}
  \caption{\footnotesize (a.) Single measurement results for the coating absorption of mirror $M_1$ in p-pol (upper, green circles: \unit[484]{mW} input light power, upper light blue dots \unit[840]{mW} input light power) and s-polarization (lower, dark blue dots): Several measurements were performed for each polarization varying the scan frequency and in p-polarization additionally the input power. The solid lines show the mean values and the dashed lines the standard deviation of the individual results. (b.) Results of the single measurements for the cavity round-trip loss (($1\!-\!\tilde{R}_{2}$)$-$($1\!-\!R_1$)) in p-polarization: At each of the cavity input powers \unit[1.5]{mW} ($\blacklozenge$), \unit[12]{mW} ($\blacktriangle$), \unit[113]{mW} ($\bullet$), \unit[484]{mW} ($\blacktriangledown$) and \unit[840]{mW} ($\blacksquare$) several single measurements were performed.}
  \label{fig:alpha}
\end{figure}

A pair of two deformed resonance peaks -- one for shortening and one for lengthening the cavity -- allowed us to obtain the quantities $R_1$, $\tilde{R}_2$ and $\alpha^{\rm coat}$. $R_1$ is the power reflectivity of in-coupling mirror $M_1$, whereas $\tilde{R}_2$ is the {\it{effective}} reflectivity of $M_2$ that includes all cavity round-trip losses apart from the transmission of $M_1$, and $\alpha^{\rm coat}$ is the absorption of the coating of $M_1$. A measurement of such a pair of resonance peaks we shall call in the following a \textit{single} measurement. To reduce errors due to statistical fluctuations, several single measurements were performed. For every single measurement a Nelder-Mead algorithm varied $R_1$, $\tilde{R}_2$ and $\alpha^{\rm coat}$, and thus minimized the deviation between simulated and measured data. The single measurements were carried out with different scan velocities to exclude systematic effects. For p-polarization, in addition to the scan frequency also the input power was varied to exclude intensity-dependent effects that were found in c-Si~\cite{Degallaix12,Khalaidovski13}. The results for our single measurements are shown in Fig.~\ref{fig:alpha}\,(a.). The upper light-blue dots show results at an input power of \unit[484]{mW}, the green circles show the results taken at an input power of \unit[840]{mW}. Further, the round-trip loss was measured at input powers of \unit[1.5]{mW} ($\blacklozenge$), \unit[12]{mW} ($\blacktriangle$), \unit[113]{mW} ($\bullet$), \unit[484]{mW} ($\blacktriangledown$) and \unit[840]{mW} ($\blacksquare$). For p-polarization, the result is $\alpha^{\rm coat}_{\rm p}=\unit[(1428 \pm 97)]{ppm}$. The results for the reflectivities are $1-R_{1, \rm p}=\unit[(4717 \pm 212)]{ppm}$ and $1-\tilde{R}_{2,\rm p}=\unit[(7882 \pm 251)]{ppm}$. 
Since in p-polarization no power dependence occurred, in s-polarization the input power was kept constant at \unit[367]{mW}, while the scan frequency was varied between \unit[5]{Hz} and \unit[500]{Hz}. Figure~\ref{fig:SiSiO2_Peaks}\,(a.)$\:$shows resonance peaks (external lengthening: red dots -- measurement, orange line -- simulation; external shortening: blue dots -- measurement, light-blue line -- simulation) detected at a scan frequency of \unit[90]{Hz}, where the two peaks for an external lengthening and shortening of the cavity were identical. A frequency of about \unit[15]{Hz} was the limit at which the thermal effect was sufficient for fitting $\alpha^{\rm coat}$. The results of the single measurements for s-pol are presented in Fig.~\ref{fig:alpha}\,(a.) (lower, dark-blue dots). The mean value (lower, purple line) and standard deviation (lower, purple dashed lines) of the results from the single measurements is $\alpha^{\rm coat}_{\rm s} = \unit[(1035 \pm 42)]{ppm}$. The results for the reflectivities are $1-R_{1,\rm s}=\unit[(529 \pm 20)]{ppm}$ and $1-\tilde{R}_{2,\rm s}=\unit[(2902 \pm 103)]{ppm}$ (= round-trip loss). Our measured value of $R_{1,\rm s}$ is in perfect agreement with the design reflectivity.

In the following, the influence of potential errors in the simulation input parameters on our results is analyzed. Therefore, we simulated the influence of the individual parameters listed in Tab.~\ref{tab:parameters} on our result beginning with the material parameters, which we took from the literature. The errors of the refractive index $n$ as well as of the thermal refractive coefficient d$n$/d$T$ were found to be negligible~\cite{leviton06}. An error of the thermal expansion $a_{\rm th}$ was also negligible since d$n$/d$T$ by far dominates the phase shift of the laser beam~\cite{Stei12}. Changes of the specific heat $c$ and density $\rho$ affected the absorption result approximately linearly. A change by \unit[10]{\%} of the thermal conductivity $k_{\rm th}$ affected the result by about \unit[4]{\%}.
Potential errors in the cavity geometry parameters were also found to be negligible. The mirror length $D$ and mirror radius $r$ did not affect our absorption result for $D,r \gg w_0$, which is valied for our setup. A measurement error of the round trip length $L$ of \unit[420]{mm} was estimated to be  $<\unit[1]{\%}$ and was thus negligible as well. $AOI$ and beam waist $w_0$ were based on $L$ and potential errors in these quantities were also negligible.
We thus conclude that the standard deviation of our measurements also includes potential errors of the input parameters if we assume that the errors of the parameters are statistically independent. 

%
%
\section{Discussion and Conclusion}

For our dielectric Si/SiO$_2$ coating produced with ion plating, absorption values of $\alpha^{\rm coat}_{\rm s}=\unit[(1035 \pm 42)]{ppm}$ and $\alpha^{\rm coat}_{\rm p}=\unit[(1428 \pm 97)]{ppm}$ were measured. These values are orders of magnitude higher than the optical absorption of state-of-the-art SiO$_2$/Ta$_2$O$_5$ coatings being in the ppm regime~\cite{Rempe92,Stei12}.\\
To transfer our results to an absorption coefficient for the deposited silicon layers, the layout of the actual dielectric Si/SiO$_2$ coating stack is required. Since the actual design was not at our disposal, we used a model coating stack that matched our measurement results for the reflectivities in s- and p-polarization  of $R_{\rm{meas,\,s/p}}=0.99947/0.99528$ using~\cite{lightmachinery}. $R_{\rm{sim,\,s/p}}=0.99957/0.99510$ was the best matching set that could be achieved. Figure~\ref{fig:experimental_setup}\,(b.) shows a schematic of the simulated coating stack including the thicknesses per single layer. The simulated reflectivity after each double layer is given on the right side in Fig.~\ref{fig:experimental_setup}\,(b.). Since the individual double layers transmit only a fraction of the laser power, the absorption coefficient needs to be based on an effective penetration depth, here given in units of double layers
\begin{eqnarray*}
D_{\rm eff,s}&=&1+(1-0.70778)+(1-0.93993)+(1-0.98832)\\
&& \quad +(1-0.99775)+(1-0.99957)=1.36665\, ,
\end{eqnarray*}
where the numbers are taken from Fig.~\ref{fig:experimental_setup} for s-polarized light. As the result, effectively 1.37 double layers are transmitted twice by \unit[100]{\%} of the input power. Assuming the absorption in the SiO$_2$ layers to be negligible \cite{Rempe92,Stei12}, this results in an absorption coefficient of $\unit[1035]{ppm}/(2\times 1.37 \times 1.02 \times \unit[101]{nm}) = \unit[(37 \pm 2)]{/cm}$ for the silicon layers. The factor 1.02 is a correction of the layer thickness due to the AOI differing from $0^\circ$. Correspondingly, for p-polarization the absorption coefficient is $\unit[(39 \pm 3)]{/cm}$. Both results are in good mutual agreement.

Our absorption coefficient of $\alpha_{\rm{si}}$(1550nm)\,$\approx \unit[40]{/cm}$ found for the ion plated silicon layers of our coating stack is lower than values of \unit[100]{/cm} up to \unit[2000]{/cm} reported for a-Si~\cite{Chittick70,Brodsky69,Loveland73}, while for c-Si the absorption coefficient is only in the order of a few ppm/cm up to a few hundreds ppm/cm at \unit[1550]{nm}~\cite{Keevers1995,Degallaix12,Stei13}.  Our result reveals a rather modest reduction of optical absorption for Si layers deposited by ion plating instead of the ion beam sputtering technique, and the coating investigated has an absorption that is still far too high for test mass mirrors in GWDs.

%
\section*{Acknowledgements}

We acknowledge support from the SFB/Transregio 7, the International Max Planck Research School (IMPRS) on Gravitational Wave Astronomy, and from QUEST, the Centre for Quantum Engineering and Space-Time Research.
%
%

\end{document}